\documentclass[runningheads,a4paper]{llncs}

\usepackage{amssymb}
\usepackage{graphicx}
\usepackage{multirow}
\usepackage{epsfig}
\usepackage{url}
\newcommand{\keywords}[1]{\par\addvspace\baselineskip
\noindent\keywordname\enspace\ignorespaces#1}

\begin{document}

\mainmatter  

\title{Towards Air Quality Estimation Using Collected Multimodal Environmental Data}

\author{Anastasia Moumtzidou\inst{1} \and Symeon Papadopoulos\inst{1} \and Stefanos Vrochidis\inst{1} \and Ioannis Kompatsiaris\inst{1} \and Konstantinos Kourtidis\inst{2} \and
George Hloupis\inst{3} \and Ilias Stavrakas\inst{3} \and Konstantina Papachristopoulou\inst{4} \and Christodoulos Keratidis\inst{4}}

\authorrunning{A. Moumtzidou et al.}

\institute{
Centre for Research \& Technology Hellas - Information Technologies Institute\\
\email{{moumtzid, papadop, stefanos, ikom}@iti.gr}
\and
Democritus University of Thrace\\ 
\email{kourtidi@env.duth.gr}
\and
Technological Education Institute of Athens\\ 
\email{hloupis@teiath.gr, ilias@ee.teiath.gr}
\and
DRAXIS Environmental Technologies Company\\ 
\email{{k.papachristopoulou, keratidis.ch}@draxis.gr}
}

\tocauthor{Anastasia Moumtzidou, Symeon Papadopoulos, Stefanos Vrochidis,  Ioannis Kompatsiaris, Konstantinos Kourtidis, George Hloupis, Ilias Stavrakas, Konstantina Papachristopoulou, Christodoulos Keratidis}
\maketitle

\begin{abstract}
This paper presents an open platform, which collects multimodal environmental data related to air quality from several sources including official open sources, social media and citizens. Collecting and fusing different sources of air quality data into a unified air quality indicator is a highly challenging problem, leveraging recent advances in image analysis, open hardware, machine learning and data fusion and is expected to result in increased geographical coverage and temporal granularity of air quality data.
\end{abstract}

\section{Introduction}

Environmental data is very important for human life and the environment. Especially, the environmental conditions related to air quality are strongly related to health issues (e.g. asthma) and to everyday life activities. Such data is measured by dedicated stations established by environmental organizations, which are usually made available through web sites and services. Furthermore, the availability of low cost hardware sensors allowed for the establishment of personal environmental stations by citizens. In parallel, the increasing popularity of social media has resulted in massive volumes of publicly available, user-generated multimodal content that can often be valuable as a sensor of real-world events \cite{Aiello2013}. This fact coupled with the rise of citizens' interest in environmental issues, has triggered the development of applications that make use of social data for collecting environmental information and creating awareness about environmental issues. 

To this end, this paper presents a new platform developed in hackAIR project\footnote{www.hackair.eu}, for gathering and fusing environmental data and specifically Particulate Matter (PM) measurements from official open sources and user generated content (including social media communities). This platform aims to contribute towards individual and collective awareness about air quality and to stimulate sustainable behaviour with respect to it.

\section{Relevant Initiatives}
There are several initiatives including projects and applications that attempt to provide citizens with environment-oriented information collected from different data sources. Table ~\ref{tab_init} contains a detailed list of such initiatives. For the sake of space, we shall briefly mention only a limited number of them: a) iSCAPE that encapsulates the concept of ‘smart cities’ by promoting the use of low cost sensors and the use of alternative solution processes to environmental problems, b) the Amsterdam Smart Citizens Lab that uses smartphones, smart watches, and wristbands, as well as open data and DIY sensors for collecting environmental data, c) AirTick, which	 estimates air quality in Singapore by analysing large numbers of photos posted in the area, and d) PESCaDO that focused on open environmental sources and provided users with personalized information.

As far as the applications are concerned, the most interesting are: a) Ubreathe that provides current and forecast air quality as well as health advice for UK, b) World Air quality that reports Air Quality Index for 500 cities around the world, c) AirForU that provides Air Quality Index, hourly updates, one day forecast, historical exposure, and personalised tips, and d) Air Visual that presents historical, real-time and forecast air quality data, including PM10, SO2, temperature using indoor and outdoor sensors.

Compared to the aforementioned initiatives, the proposed platform combines data from various sources in an effort to benefit from the reliability of open official data, the abundance and high coverage of publicly available images posted through social media, the quality and consistency of images captured by users of the platform-oriented mobile app and the reliability of the measurements of low-cost open sensor devices for relatively large numbers of community users.

\begin{table}
\caption{Relevant initiatives}
\label{tab_init}
\begin{center}
\begin{tabular}{|l|l|}
\hline
\textbf{Type}                & \textbf{Name of Initiative}                             \\ \hline
\multirow{8}{*}{Project}     & iSCAPE (\cite{iscape})                            \\ \cline{2-2} 
                             & Amsterdam Smart Citizens Lab (\cite{ascl})         \\ \cline{2-2} 
                             & PASODOBLE  (\cite{pasodoble})                     \\ \cline{2-2} 
                             & PESCaDO (\cite{pescado})                            \\ \cline{2-2} 
                             & CITI-SENSE (\cite{citisense})                       \\ \cline{2-2} 
                             & Plume Labs (\cite{plumelabs})                        \\ \hline
\multirow{7}{*}{Application} & Ubreathe (\cite{ubreathe})                           \\ \cline{2-2} 
                             & World Air quality (\cite{waq})                       \\ \cline{2-2} 
                             & Banshirne (\cite{banshirne})                          \\ \cline{2-2} 
                             & AirForU (\cite{afu})                                  \\ \cline{2-2} 
                             & Clean Air Nation (\cite{can})                         \\ \cline{2-2} 
                             & Air Visual  (\cite{av})                               \\ \hline
\end{tabular}
\end{center}
\end{table}

Apart from the aforementioned projects and applications, a work that is related to the proposed system is that of Tang et al. \cite{Tang2015} that uses the EventShop software, which provides a generic infrastructure for analyzing heterogeneous spatio-temporal data streams regarding real-world events. However, EventShop focuses mainly on the fusion, and analysis of measurements coming from different sources and is not concerned with the retrieval and analysis of multimodal data.

\section{System Architecture}
The proposed system will collect PM measurements from various sources that will be processed according to their type (i.e. text, image). The sources that are foreseen are: a) web-based official sources, b) image-based sources, and c) hardware-based sources. The aim of having different sources is to address the need for both reliable and large in number measurements.

As far as the web-based official sources are concerned, these involve publicly available open data found in environmental web sites, and web services. Regarding image-based sources, they include publicly available geotagged images posted through platforms such as Instagram, images captured by the users of the hackAIR mobile app and webcams. Finally, hardware-based sources involve low-cost open sensor devices assembled by citizens to monitor the concentration of PM.

\begin{figure}
	\centering
	\epsfig{file=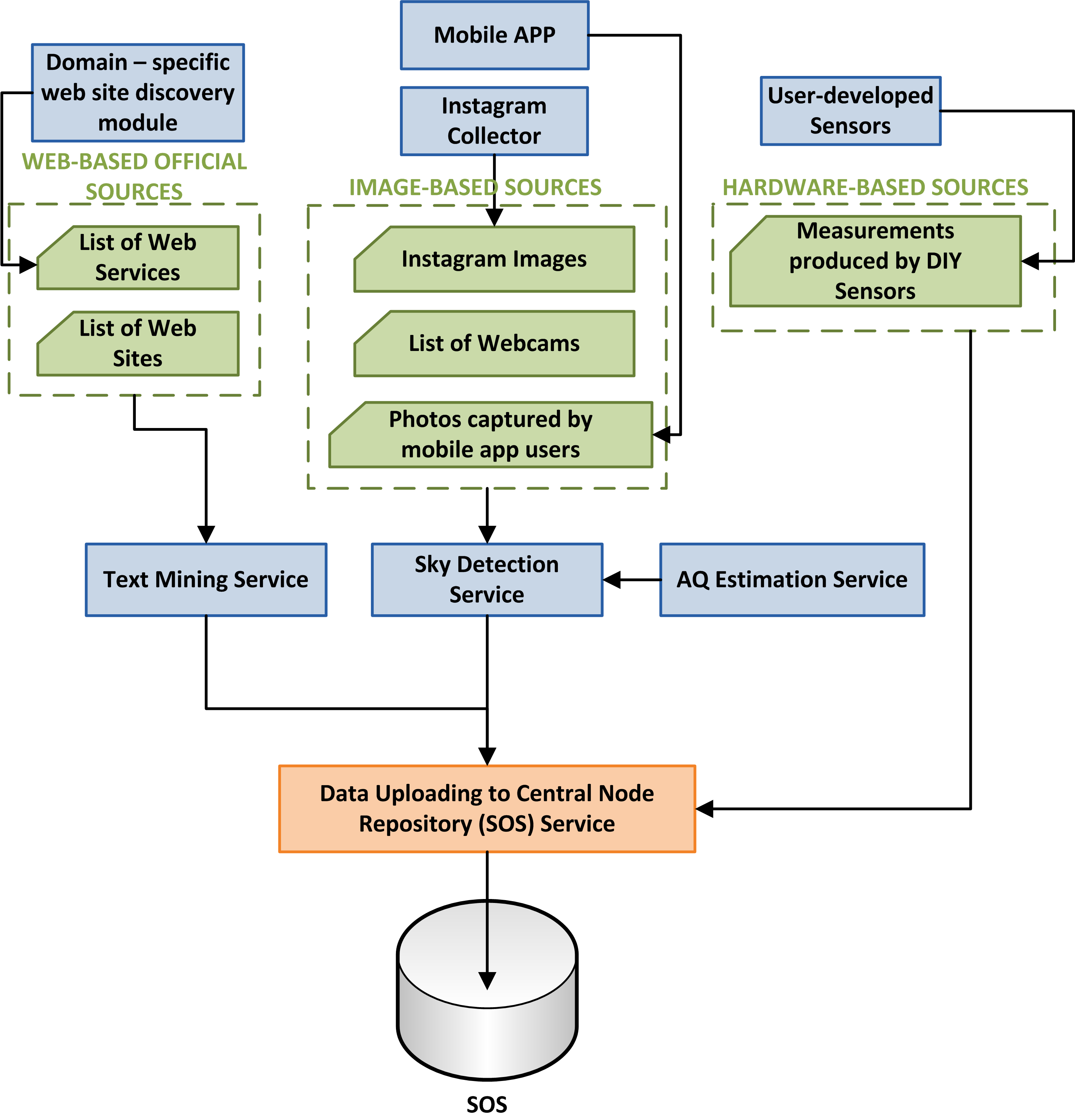, width=8cm}
	\caption{System architecture.}
	\label{fig:architecture}
\end{figure}

The diversity of sources results in obtaining multimodal in nature input data that include images, unstructured and structured text and numeric values. Depending on the type of data, different analysis procedures are foreseen. Specifically, images (coming from social media, the mobile app, and webcams) will be processed using image analysis techniques and image-based air quality estimation that provides an air quality index (e.g. low, high). In the case of web sites, the data is provided in unstructured format and thus text mining is required to extract the target information. Moreover, in the case of web services, the data is provided in structured format and thus no complex text processing is required. Finally, in the case of user-developed sensors, the sources provide numeric values. Eventually, all collected data is stored into a Sensor Observation Service (SOS) server repository \cite{Na2007}. Figure ~\ref{fig:architecture} depicts an overview of the hackAIR architecture.

In the remaining of the paper, we present the data sources used and the techniques that will be applied for retrieving data from these sources and the post-processing techniques.

\section{Data Sources \& Retrieval Techniques}
\subsection{Web-based Official Sources}
These include web services and web sites that contain environmental measurements. The discovery and indexing of web services and webcams is conducted with the help of air quality experts, while the web sites are discovered using domain specific web search and crawling, which can be divided into two main categories: the first is based on using existing general search engines to access the web and retrieve a first set of results, which are subsequently filtered with the aid of post processing techniques (\cite{Oyama2004}, \cite{Chen2001}); the second is based on using a set of predefined web sites and expanding them using focused crawling and with the help of machine learning techniques \cite{Zheng2008}.

\subsection{Image-based Data Sources}
Those include images found in: 1) media sharing platforms such as Instagram, 2) images captured from the mobile app and 3) webcams.

Regarding the images found in media sharing platforms such as Instagram, they offer the advantage of abundance and high geographic coverage. Initially, only geotagged images will be collected and thus for a known rectangle enclosing each city, a set of geo-targeting queries are submitted to the Instagram API. However, in case the number of geotagged images is not sufficient (large-scale studies on Instagram suggest that approximately 20\% of the images posted to Instagram are geotagged \cite{Manikonda2014}, additional images will be collected by retrieving images tagged with the city name or tagged with known landmarks in the city.

As far as images captured from the mobile app are concerned although they are handled in the same way as social media images, they are expected to be of much higher quality and consistency since their capturing parameters will be controlled by the hackAIR mobile app.

Finally, webcams can be used as another source of images that depict parts of the skyline of an area of interest. It should be noted that a simple preprocessing of the initial video stream is required to extract frames at specific rate.

\subsection{Hardware-based Sources}
These include air quality estimations produced by low-cost open sensor devices assembled by citizens. These estimations will be performed using open hardware and software. More specifically, the widely used Arduino development platform along with the newly introduced PSOC 4 Bluetooth Low Energy (BLE) kits will be programmed for a series of widely available PM sensors in order to enable any potential user to perform and contribute air quality measurements. Preconfigured and open software modules will be provided to users along with suggested hardware configurations in order to enable them to implement low-budget air monitoring stations. Data transmission will take place by means of BLE enabled smartphones and an associated Android application. An additional data collection system oriented to users less familiar with electronics will be used. The proposed system will be built around common off-the-shelf materials. The operation principle is to force air (using aquarium or camping pumps) to pass through a paper filter and after a predefined period of exposure to take a photo (by smartphone) of the filter. Then by using computer vision algorithms, developed specifically to handle such images, the colorization of the filter will be translated to PM estimations.
More specific, the colorization of the examined filter that will be caused by PMs will form unique or small clusters on the top layer of the filter. These clusters will be projected in image as blobs. Then a blob detector can be applied that will provide the number of identified and marked blob regions which can be correlated with PM concentration (after calibration). The algorithm has the following steps: Thresholding, grouping, merging and blob radius calculation.

\section{Data Analysis}

\subsection{Web Information Extraction}
This module involves the extraction of environmental content from web sites and web services. In case of web services the format of the data is well defined and data can be retrieved by simple JSON/XML parsing. As far as web sites are concerned, information extraction from environmental web sites cannot be realized using deep semantic analysis given that most of the information to be retrieved is not reported in text and often there is little linguistic context available. Thus, information extraction can be based on the extraction and transformation of semi-structured web content, typically in HTML format, into structured data. This task typically involves acquiring the page content, processing it, and extracting the relevant information using regular expressions. 

\subsection{Sky Detection}
The module involves two image processing operations: 1) visual concept detection based on low level feature generation and classification for detecting images that contain substantial regions of sky, and 2) localization of the sky regions within the image. As far as visual concept detection is concerned different detectors will be studied including images representation with SIFT, SURF, aggregation using VLAD and training using Logistic Regression \cite{Markatopoulou2015}. Another technique involves representation of images using a pre-trained Deep Convolutional Neural Network and training with Linear SVMs \cite{Krizhevsky2012}. As far as sky localization is concerned, the selective search technique will be tested that combines the strength of both an exhaustive search and segmentation \cite{Uijlings2013}.

\subsection{Air Quality Estimation}
The module involves the estimation of air quality from user-generated photos or webcams. In general, several studies (\cite{Zerefos2014}, \cite{Saito2015}) have shown that the color ratio R/G in digital images can be used to derive information about the aerosol content of the atmosphere. Specifically, the system will use the Santa Barbara DISTORT Atmospheric Radiative Transfer Model (SBDART, \cite{Ricchiazzi1998}) to simulate the R/G ratios for a set of solar zenith angles (SZA) and a set of Aerosol Optical Depths (AOD), the latter approximating the PM load. The R/G ratio will be approximated using the ratio of the diffuse irradiance of two wavelengths (550 nm and 700 nm) rather than the radiance. The resulting R/G ratios will be used to create a 3-D lookup table (Table RG) containing R/G, AOD and SZA. Further, the SZA for each day of the year and each hour of the day will be computed for the geographical latitude/longitude of the urban areas of interest and 3-D lookup tables will be created for each urban area of interest containing SZA, Day of Year (DoY) and Time of Day (ToD). Each geotagged image whose coordinates are within the area of interest will be processed for extraction of information on the mean sky R/G ratio that will be computed automatically for the image, the image coordinates, the DoY and ToD the image was taken.

The table for the respective coordinates will be accessed, receiving as input the DoY and ToD and giving as output the SZA. The Table RG will be accessed, receiving as input the SZA and the image R/G ratio and giving as output the AOD, which, together with the image coordinates will be used to plot the AOD value on a city map.

\section{Data Storage and Indexing: SOS Repository}
In order to store and index efficiently the information retrieved from the previously described sources, it is essential to employ a database, which can store efficiently the measurements along with the related information (i.e. date-time, area coverage, and source). Thus, each source can be considered as a sensor, which provides measurements. Hence, a natural option for handling this information is through a Sensor Observation Service infrastructure, which provides a generic and flexible means for accessing data produced by sensors \cite{Na2007}. This includes access to measurements of the sensors, as well as access to information about the observed features of interest and information about the sensor. The flexibility of the Observation and Reference Model (O\&M) can be used for accessing heterogeneous data via a single standard service interface \cite{Cox2011}.

\section{Conclusions}
The hackAIR system builds upon the concept of monitoring and fusing heterogeneous and user-generated air quality monitoring resources towards providing reliable measurements \cite{Epitropou2011}. Towards fusing observational data from the aforementioned sources we plan to evaluate methods based on geostatistics, which build upon previous studies demonstrating its feasibility, such as \cite{Denby2008}. In \cite{Denby2008}, residual kriging is used to combine the sensor observations with a static base map obtained from a geophysical or statistical model. 
Other fusion techniques that we plan to evaluate include combination of land-use regression techniques with statistical air quality modelling \cite{Johansson2015}.
Eventually the fused data will be used to provide personalised services with respect to environmental issues that will raise the awareness of the citizens on air quality and engage them actively in measuring and publishing air pollution levels.

\subsubsection*{Acknowledgments.} This work is partially funded by the European Commission under the contract number H2020-688363 hackAIR.

\end{document}